\documentclass[prd,twocolumn,amsmath]{revtex4}
\usepackage{amssymb,amsmath}
\usepackage{graphicx}
\usepackage{color}

\usepackage{soul}
\usepackage[dvipsnames]{xcolor}

\begin{document}
\title{Quantum Scattering on a Cone revisited}

\author{V. S. Barroso}
\email[]{barrosov@ifi.unicamp.br}
\affiliation{Instituto de F\'isica ``Gleb Wataghin'', Universidade Estadual de Campinas, 13083-859, Campinas, S\~ao Paulo, Brazil}
\author{J. P. M. Pitelli}
\email[]{pitelli@ime.unicamp.br}
\affiliation{Departamento de Matem\'atica Aplicada, Universidade Estadual de Campinas, 13083-859, Campinas, S\~ao Paulo, Brazil}

\begin{abstract}

We revisit the scattering of quantum test particles on the conical $(2+1)$-dimensional spacetime and find  the scatteting amplitude as a function of the boundary conditions  imposed at the appex of the cone. We show that the boundary condition is responsible for a  purely analytical term in the scattering amplitude, in addition to those  coming from purely topological effects. Since it is possible to have non-equivalent physical evolutions for the wave packet (each one corresponding to a different boundary condition), it seems  crucial to have an observable quantity specifying which evolution has been prescribed.

\end{abstract}

\maketitle

\section {Introduction}

In classically  singular spacetimes, the evolution of wave packets may not be uniquely determined by the initial conditions. It is possible to have an infinite number of boundary conditions at the classical singular point, each one giving a reasonable physical evolution. To predict physical effects in such spacetimes, we need to know which evolution has been  prescribed. In this way, it is essential to have an observable quantity depending on the possible boundary conditions. This is the main goal of this paper. We will show that the differential cross section of wave scattering on the cone carries the information we need.

In globally hyperbolic spacetimes, the propagation of particles and waves are uniquely determined. Given the initial position and velocity of a particle, the classical trajectory can be extended for all times. In a similar way, given the initial wave packet $\Psi(0)$ [and possibly $\dot{\Psi}(0)$] on a Cauchy surface, there is a well defined evolution $\Psi(t)$. However, in  a singular spacetime, a classical trajectory which reaches the singularity can not be extended and the future of the particle becomes unkown. In a similar way, since no Cauchy surface exists, the evolution of waves may be ambiguous.

In static singular spacetimes~\cite{pitelli1}, a boundary condition must be imposed at the singular point in order to find the evolution of the wave packet.  These boundary conditions are the ones which turn into self-adjoint the spatial part of the wave operator, giving rise to a sensible dynamics~\cite{wald1}. If there is only one boundary condition, there is no ambiguity and we say that the spacetime is quantum mechanically (QM) non-singular~\cite{horowitz}  and that the singularity has been ``healed'' by quantum mechanics. On the other hand,  if there are an infinite number of possible boundary conditions, the evolution is uncertain. Since there is no priviledged evolution, we say that the spacetime in this case is QM singular~\cite{horowitz}. The $(2+1)$-dimensional cone is an example of a QM singular spacetime and will be used as a toy model to show that it is possible to find observational parameters related to boundary conditions in Nature.

It is well known that solutions of  Einstein  field equations
\begin{equation}
R_{\mu\nu}=\kappa\left[T_{\mu\nu}-g_{\mu\nu}T^{\lambda}_{\phantom{\lambda}\lambda}\right]
\end{equation}
in $(2+1)$-dimensions are locally flat in the absence of matter. This happens essentially because the Riemann curvature tensor may be written as
\begin{equation}\begin{aligned}
R_{\alpha\beta\gamma\delta}=&g_{\alpha\gamma}R_{\beta\delta}-g_{\alpha\delta}R_{\beta\gamma}-g_{\beta\gamma}R_{\alpha\delta}+g_{\beta\delta}R_{\alpha\gamma}\\&-\frac{1}{2}\left(g_{\alpha\gamma}g_{\beta\delta}-g_{\alpha\delta}g_{\beta\gamma}\right)R
\end{aligned}
\end{equation}
and $T_{\mu\nu}=0$ implies $R_{\mu\nu}=0$ through Einstein equations~\cite{gott}.
In addition to the $(2+1)$-dimensional Minkowski solution, there is a solution with nontrivial topology, which represents a massive point object and is identified as the product of a time-like straight line and a two-dimensional  cone. It has the following metric~\cite{staruszkiewicz}
\begin{equation}
    \label{eq:cone}
    ds^2=-dt^2+dr^2+\alpha^2r^2d\theta^2,
\end{equation}
with $0< r < \infty$, $0\leq \theta \leq 2\pi$ and $0<\alpha< 1$. The mass $M$ of the object  located at $r=0$ is related to $\alpha$ by $M=2\pi(1-\alpha)/\kappa$. The cone generated by the spatial part of metric \eqref{eq:cone} has opening angle given by $\varphi=2\sin^{-1}\alpha$ and in three dimensions is parametrized by $z(r)=(\alpha^{-2}-1)^{1/2}r$. It has its vertex at $r=0$ which is a classical spacetime singularity.

Following the definition of quantum singularities due to Horowitz and Marolf~\cite{horowitz} (see also~\cite{wald2}), it is  known that the cone is also quantum mehanically singular, i.e., there are an infinite number of possible boundary conditions  at the appex of the cone with sensible dynamics. In Ref.~\cite{kay}, these boundary conditions have been found. However, the scattering of waves on the cone has only been studied using a particular boundary condition~\cite{deser}, namely, the Friedrichs boundary condition~\cite{reed}.

This gives rise to the following question: how is the dynamics of a quantum test particle affected by the singularity at the cone vertex if we consider an arbitrary boundary condition? We attempt to answer this question by analysing the scattering behavior of a scalar field on the cone. As we will see, the contribution of a general boundary condition to the differential cross section is purely analytical, in the sense that it is always present and is independent of the angular deficit $\Delta=2\pi(1-\alpha^2)$. This contribution also adds up to the purely topological contribution which depends directly on $\alpha$.

This paper is organized as follows: in Sec. \ref{secii} we give a brief review on the theory describing the dynamics of quantum test particles in classically singular spacetimes. Then, in Sec. \ref{seciii}, we present a solution for Klein-Gordon equation on the cone with the appropriate boundary conditions at $r=0$. We revisit the scattering of a quantum test particle on a conical spatial background in Sec. \ref{seciv}. In Sec. \ref{conclusions} we conclude by discussing the differences between our results and previous ones.

\section{Quantum singularities and self-adjoint operators}
\label{secii}

A classical singularity is indicated by incomplete geodesics or incomplete curves of bounded acceleration~\cite{geroch}. Accordingly, the evolution of classical particles following these geodesics may not be defined for all values of its affine parameter~\cite{hawking}. At the center of a spherically symmetric black hole, for example,  we have a very strong singularity, with infinite tidal forces. However, it is also possible to have milder singularities as solutions of Einstein equations. So is the case of the cosmic string background~\cite{vilenkin}, given by the metric 
\begin{equation}
    \label{eq:cosmic strings}
    ds^2=-dt^2+dr^2+\alpha^2r^2d\theta^2+dz^2,
\end{equation}
which is locally flat (each section $z=\textrm{constant}$ is a cone). In this way, there are geodesics which approaches the singularity at $r=0$ feeling zero tidal forces. This  is an example of a naked singularity.

In some cases, a naked singularity can be healed when the spacetime is probed by waves. As an example, we have the hydrogen atom, in which the position of the proton ($r=0$) is a  classical singularity. However, when solving Schr\"odinger equation, we must only impose square-integrability to find a complete set of orthonormal solutions. In this way, the evolution $\Psi(t)$ of any wave-packet is uniquely determined by the initial condition $\Psi(0)$.  Since there is no ambiguity in the solution of the wave equation for the hydrogen atom, we say that it is QM non-singular. As the evolution of waves is unique in QM non-singular spacetimes, physical predictions are then uniquely determined. 

In a QM singular spacetime, since the evolution of waves is no longer unique, the physical system does not give unique physical predictions. Each physical evolution is attached with a specific boundary condition at the singularity. We present the general theory of QM singularities due to Horowitz and Marolf~\cite{horowitz} in what follows.

Let us restrict to static spacetimes with timelike Killing vector $\xi^\mu$, where $t$ denotes its parameter. Klein-Gordon equation 
\begin{equation}
    (\nabla_\mu\nabla^\mu-\mu^2)\psi=0,
    \label{eq:kg}
\end{equation}
can be rewritten as
\begin{equation}
    \label{eq:wavekg}
    \frac{\partial^2\psi}{\partial t^2}=-A\psi,
\end{equation}
where $A\equiv -VD^i(VD_i)+\mu^2$, $V^2=-\xi^\mu\xi_\mu$ and $D_j$ is the spatial covariant derivative on a static slice of space $\Sigma$.

Since we do not know what exactly happens at the singularity, consider the domain of operator $A$ as being $C_{0}^{\infty}(\Sigma)$. Since the singular points are not part of $\Sigma$, the singularity is not being considered. With this choice, it is easy to see that the operator $\left(A,C_{0}^{\infty}(\Sigma)\right)$ is symmetric and positive definite but not self-adjoint. However, it has at least one self-adjoint extension (Friedrichs extension~\cite{reed}).

A general solution for Eq. (\ref{eq:wavekg}) has the form
\begin{equation}
    \psi_E(t)=\psi(0)\cos(A_E^{1/2}t)+\dot{\psi}(0)A_E^{-1/2}\sin(A_E^{1/2}t),
\end{equation}
where $A_E$ is a self-adjoint extension of the operator $A$. If this extension is unique, $A_E$ represents a single operator ($A$ is essentially self-adjoint) and, since there is no ambiguity in the evolution of a wave packet, the spacetime is said to be QM non-singular. If there are an infinite number of self-adjoint extensions, $E$ represents a parameter and the spacetime is QM singular. To each self-adjoint extensions, there corresponds a boundary condition at the singularity. In the next section, we give an example of this procedure for the Kleing-Gordon equation on the cone. Following Ref.~\cite{kay}, we present each boundary condition which turns into self-adjoint the spatial part of the wave operator.

\section{Klein-Gordon equation on the cone}
\label{seciii}

Klein-Gordon equation  on the conical $(2+1)$-dimensional spacetime given by the metric \eqref{eq:cone} is written as
\begin{align}
    \frac{\partial^2\phi(t,r,\theta)}{\partial t^2}&=\left(\frac{\partial^2}{\partial r^2}+\frac{1}{r}\frac{\partial}{\partial r}+\frac{1}{\alpha^2r^2}\frac{\partial^2}{\partial \theta^2}-\mu^2\right)\phi(t,r,\theta) \nonumber\\
    &\equiv (\Delta -\mu^2)\phi(t,r,\theta).
    \label{eq:kgcone}
\end{align}
Solutions may be separated as $\phi(t,\vec{r})=e^{-i\omega t}\Psi(r,\theta)$ and Eq. \eqref{eq:kgcone} reduces to
\begin{equation}
    -\Delta \Psi(r,\theta)=(\omega^2-\mu^2)\Psi(r,\theta).
\end{equation}

In section \ref{secii}, we discussed the importance of having $A$ as essentially self-adjoint to ensure uniqueness of the solution. However, in the conical spacetime,  this is not the case. As discussed in~\cite{kay}, the operator $-\Delta$ on the domain $C_0^\infty(\mathbb{R}^+\times \mathbb{S})\subset \mathcal{L}^2(\mathbb{R}^+\times \mathbb{S},rdrd\theta)$ has a family of self-adjoint extensions $-\Delta^R$ parametrized by one parameter $R\in[0,\infty)$. If we consider another separation of variables to the solution, namely $\Psi(r,\theta)=\sum_m f_m(r)e^{im\theta}$, our operator $-\Delta$ reduces to
\begin{equation}
    -\Delta_m=-\frac{\partial^2}{\partial r^2}-\frac{1}{r}\frac{\partial}{\partial r}+\frac{m^2}{\alpha^2r^2}.
\end{equation}
These $-\Delta_m$ on the domain $C_0^\infty(\mathbb{R}^+)\subset \mathcal{L}^2(\mathbb{R}^+,rdr)$ are essentially self-adjoint for $m\neq0$. Nevertheless, for $m=0$, $-\Delta_0$ has infinitely many self-adjoint extensions, $\{-\Delta_0^R,R\in[0,\infty)\}$. As previouslly presented, to every extension $-\Delta_0^R$ there must be an associated boundary condition at the singularity ($r=0$), as follows~\cite{kay}
\begin{align}
    \lim_{r\rightarrow 0}\left[\ln \left(\frac{r}{R}\right)r\frac{d}{dr}-1\right]f_0(r)=0, \quad &\text{for} \ R\in(0,\infty), \\
    \lim_{r\rightarrow0}r\frac{d}{dr}f_0(r)=0, \quad &\text{for} \ R=0.
\end{align}

As one solves the eigenvalue problem $-\Delta_mf_m(r)=\lambda f_m(r)$ with the appropriate boundary conditions, we find that, for $m\neq0$ and $m=0$ with $R=0$, there is a complete set of eigenfunctions with positive eigenvalues $k^2$. For $m=0$ and $R\neq0$, $-\Delta_0^R$ may be negative. If we redefine the boundary condition parameter as 
\begin{equation}
    q=2e^{-\gamma}R^{-1}, \ (\gamma=\text{Euler-Mascheroni constant})
\end{equation}
the operator $-\Delta_0^R+\mu^2$ has a negative eigenvalue $-\omega_q^2=-q^2+\mu^2$ as long as $q>\mu$. In this case, it is possible to have a solution of the form
\begin{equation}
\Psi(t,r)=K_0(\omega_q r)e^{\pm \omega_q t}.
\label{unstable}
\end{equation}
The positive exponential leads to an unstable configuration if the wave equation appears as a linear perturbation of the spacetime. Since physical predictions in unstable spacetimes are meaningless, we will consider the case $0\leq q\leq \mu$. Then we have only positive eigenvalues for the operator  $-\Delta_0^R+\mu^2$ and the complete set of solutions of Klein-Gordon equation is given by
\begin{equation}
\left\{\frac{1}{\sqrt{2\pi}}\frac{J_0(kr)+\beta(k)N_0(kr)}{\sqrt{1+\beta^2(k)}}\right\}\bigcup \left\{\bigcup_{m\neq 0}{\frac{1}{\sqrt{2\pi}}J_{\frac{|m|}{\alpha}}(kr)}\right\}
   \label{eq:solkg}    
\end{equation}
where $J_n$ is the $n$-th order Bessel function, $N_0$ is the $0$-th order Neumann function and $\omega_k^2=k^2+\mu^2$, $m\in\mathbb{Z}- \{0\}$ and 
\begin{equation}
    \beta(k)=\frac{\pi}{2}\left[\ln\left(\frac{q}{k}\right)\right]^{-1}.
\end{equation}
Note that $q=0$ corresponds to $\beta=0$, so  that in this case only regular solutions at $r=0$ are considered (Friedrichs boundary condition). 

We point out that this development is completely applicable to the nonrelativistic case by simply setting $k=\sqrt{2\mu \omega_k}/\alpha$. It is now clear that solution \eqref{eq:solkg} is not unique, for it depends on the chosen boundary condition. Therefore, the conical (2+1)-dimensional spacetime is quantum mechanically singular when tested by a Klein-Gordon field.

\section{Quantum scattering revisited}
\label{seciv}

In~\cite{deser}, Deser and Jackiw studied quantum scattering on the cone. We revisit their work in a relativistic version, considering now solution \eqref{eq:solkg}, which takes into account the appropriate boundary conditions at the vertex. Bound states are not relevant in scattering, so the spatial part is considered as
\begin{align}
    \Psi(r,\theta)=&\frac{a_0}{\sqrt{2\pi}}\frac{J_0(kr)+\beta(k)N_0(kr)}{\sqrt{1+\beta^2(k)}} \nonumber\\ &\qquad+\sum_{m\neq0}\frac{a_m}{\sqrt{2\pi}} J_{\frac{|m|}{\alpha}}(kr)e^{im\theta}.\label{eq:solcone}
\end{align}

We follow the procedure in~\cite{deser}, so the total wave is:
\begin{equation}
    \psi(r,\theta)=\psi_{in}(r,\theta)+\psi_{sc}(r,\theta), \label{eq:sep1}
\end{equation}
where $\psi_{in}$ and $\psi_{sc}$ are the incident and the scattered waves, respectively. Both satisfy the following asymptotic behavior as $r\rightarrow\infty$
\begin{align}
    \psi_{in}(r,\theta)& \sim e^{ikr\cos\theta},\\
    \psi_{sc}(r,\theta)& \sim\sqrt{\frac{i}{r}}f(\theta) e^{ikr}. 
\end{align}
In order to compare both asymptotic forms of $\Psi$ and $\psi$, we must use the following relation
\begin{equation}
    e^{ikr\cos\theta}=\sum_{m=-\infty}^\infty e^{im\frac{\pi}{2}}J_m(kr)e^{im\theta},
\end{equation}
and the asymptotic forms of Bessel and Neumann functions
\begin{align}
    J_m(kr)&\sim \sqrt{\frac{2}{\pi kr}}\cos\left(kr-m\frac{\pi}{2}-\frac{\pi}{4}\right), \\
    N_m(kr)& \sim \sqrt{\frac{2}{\pi kr}}\sin\left(kr-m\frac{\pi}{2}-\frac{\pi}{4}\right).
\end{align}

When comparing the asymptotic forms of equations \eqref{eq:solcone} and \eqref{eq:sep1}, from incident modes $e^{-ikr}$ we have
\begin{align}
    a_0&=\sqrt{2\pi}\frac{1-i\beta(k)}{\sqrt{1+\beta(k)^2}}, \\
    a_m&=\sqrt{2\pi}e^{-i\frac{|m|}{2}(\omega_c-\pi)}.
\end{align}
From scattered modes $e^{ikr}$ we get the scattering amplitude
\begin{align}
    f(\theta)=\frac{1}{\sqrt{2\pi k}}&\left\{\frac{-2\beta(k)[1-i\beta(k)]}{1+\beta(k)^2}\right.\nonumber\\
    &\qquad\left.-i\sum_{m=-\infty}^{\infty}(e^{-i|m|\omega_c}-1)e^{im\theta}\right\},\label{eq:scamp1}
\end{align}
where $\omega_c\equiv\pi(\alpha^{-1}-1)$ is the angle between the projections of the asymptotic paths of a classical particle onto the x-y plane (see~\cite{deser}). 

Now the total wave becomes
\begin{align}
    \Psi(r,\theta)=&\frac{-\beta(k)[1-i\beta(k)]}{1+\beta(k)^2}\{iJ_0(kr)-N_0(kr)\}\nonumber \\
    &\qquad+\sum_{m=-\infty}^{\infty}e^{-i\frac{|m|}{2}(\omega_c-\pi)}J_{\frac{|m|}{\alpha}}(kr)e^{im\theta},\label{eq:totwave}
\end{align}
Note the appeareance of $\beta$, which is related to the choice of the boundary condition. It does not depend on $\alpha$ and adds up to the purely topological terms. This term represents a point interaction between the incoming wave and the appex of the cone.

Equation \eqref{eq:scamp1} may be rewritten after a few regularizations as 
\begin{align}
     \sqrt{2\pi k}f(\theta)&=\frac{-2\beta(k)[1-i\beta(k)]}{1+\beta(k)^2}+\frac{\sin\omega_c}{\cos\omega_c-\cos\theta} \nonumber \\
        &-i\pi\sum_n\Big(\delta(\theta+\omega_c-2\pi n)+\delta(\theta-\omega_c-2\pi n)\nonumber\\
        &\hspace{50pt}-2\delta(\theta-2\pi n)\Big).\label{eq:scamp2}
\end{align}

This scattering amplitude cannot satisfy the Optical Theorem, since its delta functions and divergences at $\theta=\pm\omega_c$ invalidate integration over all angles between $0$ and $2\pi$. However, one can check Klein-Gordon probability current remains divergenceless and, then, solution \eqref{eq:totwave} holds probability conservation. 

As proposed by Deser and Jackiw, the part of the scattered wave that asymptotically gives rise to deltas in $f(\theta)$ may be replaced into the incoming wave. We separate the total wave function $\psi$ arbitrarily as
\begin{equation}
    \psi(r,\theta)=\tilde{\psi}_{in}(r,\theta)+\tilde{\psi}_{sc}(r,\theta),
    \label{eq:sep2}
\end{equation}
with the following asymptotic conditions
\begin{equation}
    \tilde{\psi}_{sc}(r,\theta) \sim\sqrt{\frac{i}{r}}\tilde{f}(\theta) e^{ikr},
\end{equation}
\begin{equation}
    \tilde{f}(\theta)=\frac{1}{\sqrt{2\pi k}}\bigg\{\frac{-2\beta(k)[1-i\beta(k)]}{1+\beta(k)^2}+\frac{\sin\omega_c}{\cos\omega_c-\cos\theta}\bigg\}. \label{eq:scamp3}
\end{equation}

\begin{figure}[h]
    \centering
    \includegraphics[width=.4\textwidth]{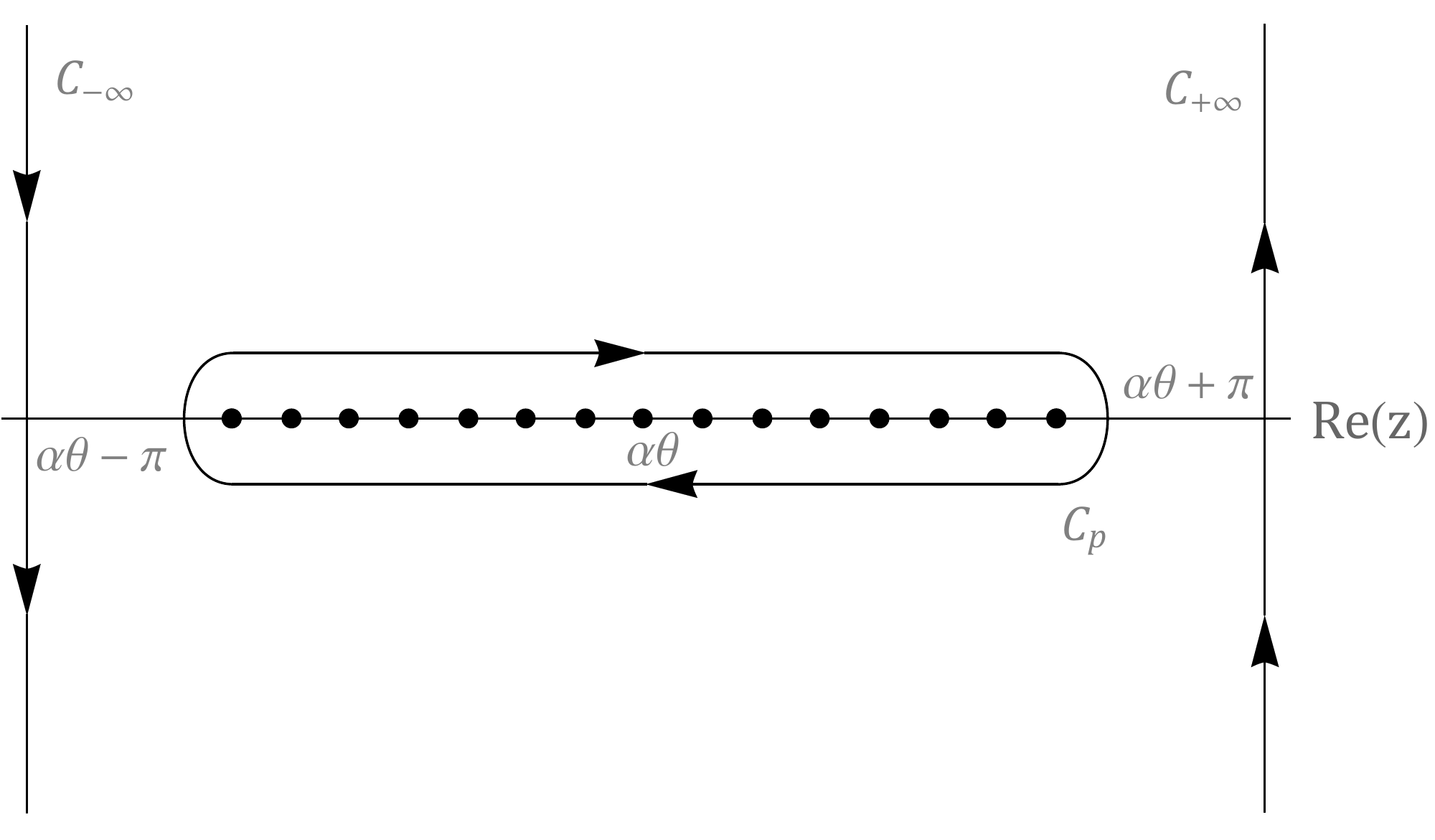}
    \caption{Integration contour $C$ for Eq. \eqref{eq:contour} separated in other three: $C_{-\infty}$, $C_{+\infty}$ and $C_p$. Contour $C_p$ is a clockwise Cauchy contour around the real poles of $\tan(z/2\alpha)$ between $\alpha\theta-\pi$ and $\alpha\theta+\pi$.}
    \label{fig:contour}
\end{figure}

Now we must find a new incident wave resulting from this redefinition. In~\cite{deser}, only the second term in equation \eqref{eq:totwave} correspond to the total wave and the authors find a contour integral representation for it. To do so, they use Schl\"afli's representation for Bessel functions, given by
\begin{equation}
    J_\nu(x)=\frac{1}{2\pi}\int_\Gamma \ dz e^{-i(k\sin z-\nu z)},
\end{equation}
where $\Gamma$ is a complex contour coming from $-\pi+i\infty$ to $-\pi$, passing by the real axis to $\pi$, and then returning to $\pi+i\infty$. Following this procedure, the sum in Eq. \eqref{eq:totwave} becomes
\begin{align}
    \sum_{m=-\infty}^{\infty}&e^{-i\frac{|m|}{2}(\omega-\pi)}J_{\frac{|m|}{\alpha}}(kr)e^{im\theta}\nonumber\\
    &=\frac{1}{4\pi i}\int_C \ dz \tan\left(\frac{z}{2\alpha}\right)e^{-ikr\cos(z-\alpha\theta)}\equiv\frac{1}{4\pi i}I_C, \label{eq:contour}
\end{align}
where $C$ is the contour given in Fig. \ref{fig:contour}.

Integration over $C_{+\infty}$ and $C_{-\infty}$ can be rewritten using the function
\begin{equation}
    \chi(r,\xi)=\int_{-\infty}^{\infty}dy\ e^{ikr\cosh y}\tan\left(\xi+i\frac{y}{2\alpha}\right).
\end{equation}
Cauchy's residue theorem allows us to express integration over $C_p$ as the sum of all residues of the integrated function at the poles. 
Finally, integral $I_C$ can be separated in three others over the contours presented in Fig. \ref{fig:contour}, as follows
\begin{align}
    I_C &=\left(\int_{C_{+\infty}}+\int_{C_{-\infty}}+\int_{C_p}\right)dz\tan\left(\frac{z}{2\alpha}\right)e^{-ikr\cos(z-\alpha\theta)} \nonumber\\
    &=i\left[ \chi\left(r,\frac{\theta}{2}+\frac{\pi}{2\alpha}\right)-\chi\left(r,\frac{\theta}{2}-\frac{\pi}{2\alpha}\right) \right]\nonumber\\
    &\qquad \qquad+4\pi i \alpha \sum_{\substack{n\\\alpha|\theta_n|<\pi}}e^{-ikr\cos(\alpha\theta_n)}, \label{eq:ic}
\end{align}
where $\theta_n\equiv\theta-(2n+1)\pi$.

We compared the asymptotic forms of \eqref{eq:totwave} and \eqref{eq:sep2}, as in \cite{deser}, and by their asymptotic contributions identified our new incoming and scattered waves

\begin{widetext}
    \begin{equation}
        \Psi(r,\theta)=\underbrace{\alpha \sum_{\substack{n\\\alpha|\theta_n|<\pi}}e^{-ikr\cos(\alpha\theta_n)}}_{\tilde{\psi}_{in}(r,\theta)}+\underbrace{\frac{-\beta(k)[1-i\beta(k)]}{1+\beta(k)^2}\{iJ_0(kr)-N_0(kr)\}+\frac{1}{4\pi}\left[ \chi\left(r,\frac{\theta}{2}+\frac{\pi}{2\alpha}\right)-\chi\left(r,\frac{\theta}{2}-\frac{\pi}{2\alpha}\right) \right]}_{\tilde{\psi}_{sc}(r,\theta)}. \label{eq:totwavenew}
    \end{equation}
\end{widetext}

As a cone is not an asymptotically flat spacetime, topological scattering may produce those undesirable deltas at the amplitude \eqref{eq:scamp2}. All these delta functions had their contributions placed into the incident wave from Eq. \eqref{eq:totwavenew}. This new incident wave can be seen as a composition of plane waves in variate directions depending on the deficit angle $\alpha$ and carrying the topological characteristics of space. We treat it as a redefinition of plane waves on the cone, for the original definition led to the appearance of deltas.  

The incident wave we found in \eqref{eq:totwavenew} is the same as Deser and Jackiw found in~\cite{deser}, this shows the prescription of a boundary condition at the vertex affects only the scattered wave. Furthermore, the topological scattering is responsible for the second term of $\tilde{\psi}_{sc}$ in \eqref{eq:totwavenew} and was already found by in~\cite{deser}. We point out that the term in $\tilde{\psi}_{sc}$ results from a point-like interaction of the wave and the localized massive object at the vertex.

In~\cite{deser}, the incident wave $\tilde{\psi}_{in}$ would be scattered by the spacetime topology, generating part of our scattered wave. In our picture, a spherically symmetric term appears at $\tilde{\psi}_{sc}$ as the incident wave perceive the boundary condition at $r=0$, showing it is a purely analytic interaction.

\begin{figure}[ht]
    \centering
    \includegraphics[width=.45\textwidth]{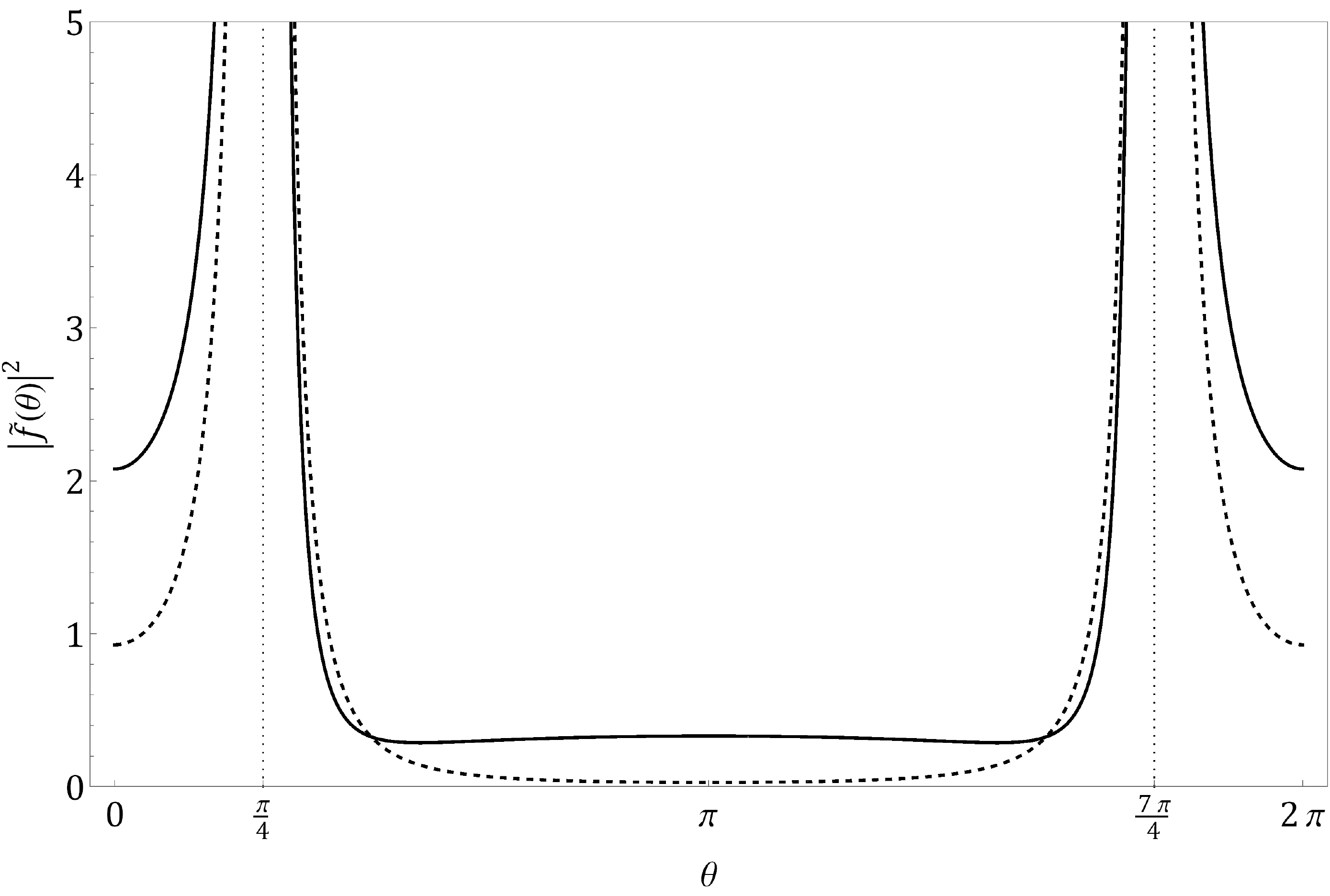}
    \caption{Plot of $|\tilde{f}(\theta)|^2$ for $\beta=0$ (dashed) and $\beta\neq 0$ (filled line). The parameters $\alpha$, related to the deficit angle  and $q$, which sets the boundary condition are $\alpha=0.8$ and $q=10$. The frequency $k$ is set equal to one. Dotted horizontal lines indicate the divergence angles $\omega_c=\pi/4$ and $\theta=2\pi-\omega_c=7\pi/4$ for the amplitude.}
    \label{fig2}
\end{figure}

In Fig.~(\ref{fig2}) we show the behavior of $|\tilde{f}(\theta)|^2$, when $\beta=0$ and there are only topological effects, and when $\beta\neq 0$ and a purely analytical term arises. There are  divergences at the classical scattering angle $\theta=\omega_c$, as well as at $\theta=2\pi-\omega_c$. These divergences appear due to topological effects as in Ref~\cite{deser} and are the signatures of the cone. The main effect of a non-null $\beta$ appears at $\theta=0$. By looking the scattering at this angle we can infer the choice of the boundary condition.

\section{Final Remarks}
\label{conclusions}
The $(2+1)$-dimensional cone was used as a toy model to illustrate the effects of an arbitrary choice of boundary conditions in QM singular spacetimes. Studying the scattering of waves, we showed that the differential cross section depends explicitly on the boundary condition, so that, through the observation of scattered waves, it may be possible to infer which evolution has been prescribed. If we want to construct quantum field theory (QFT) in non-globally hyperbolic spacetimes (see~\cite{dappiaggi} for QFT in AdS spacetime with general boundary conditions) and predict physcal observables, we need to know which evolution has been preferred by Nature. Our result gives us a hint of how to solve this question. We also argue that this simple model can be extended to more significant spacetimes, such as the spacetime of a cosmic string  and the spacetime of a global monopole~\cite{barriola}. These spacetimes are QM singular (see Refs.~\cite{konkowski,pitelli2}) so that physical observables will depend on the boundary conditions. 

To the best of our knowledge, for the first time an observable has been related to the prescribed evolution. In general, predictions in naked singular spacetimes are meaningless, since there is always an unknown parameter (the exact interaction between classical test particles and the singularity) which are not predicted by General Relativity. We showed that with the introduction of quantum mechanics we are able to find the analytical interaction between waves and the singularity by means of a single observation. Now that we can find the prescribed evolution, the next step would be the search for other observables (the expectation value of the renormalized stress tensor, for example~\cite{konkowski2}) to see how different are the predictions compared to the usually chosen Friedrichs boundary condition.

Since the perturbation of the spacetime leads to the wave equation, it is also possible that the stabillity of QM singular spacetimes depends on the physical prescription~(see discussion bellow Eq.~(\ref{unstable})). If perturbations with a wide range of possible boundary conditions are present, the spacetime wil certainly be unstable. This can explain why such spacetimes have never been observed.

\acknowledgments

The authors thank Alberto Saa and  Ricardo A. Mosna for insightful discussions and clarifying comments. The authors acknowledge support from the Sao Paulo Research Foundation (FAPESP) Grant No.~2013/09357-9. V.~S.~B. acknowledges support from CNPq Grant No.~132433/2017-6. J.~P.~M.~P.~also acknowledges support from FAPESP Grant No.~2016/07057-6 and from FAEPEX Grant No.~2693/16.

\end{document}